\documentclass[12pt]{article}

\usepackage{amsmath, amsfonts, amssymb}

\usepackage[style=authoryear, sorting=nyt,
            maxcitenames=2, mincitenames=1,
            maxbibnames=3, minbibnames=1]{biblatex}
\DeclareDelimFormat{nameyeardelim}{\addcomma\space}
\DeclareDelimFormat[finalnamedelim,bibliography]{finalnamedelim}{\addcomma\space}

\AtEveryBibitem{\ifentrytype{article}{\clearfield{title}\clearfield{number}}{}}
\DeclareFieldFormat{journaltitle}{#1}
\DeclareFieldFormat{pages}{#1}

\usepackage{graphicx}
\usepackage{indentfirst}
\usepackage{threeparttable}
\usepackage{booktabs} 
\usepackage{hyperref}
\usepackage{url}
\usepackage{xurl}
\usepackage{tabularx}

\begin{document}

\title{\LARGE \bf Contribution of coronal holes to geoeffective events of solar cycle 25}

\author{\bf O.~A.~Andreeva$^{1}$\thanks{E-mail: olga@craocrimea.ru}}

\date{\it \small $^{1}$ Crimean Astrophysical Observatory, \\
Nauchny, Bakhchisaray Region, 298409, Crimea, Russia}

\maketitle

\renewcommand{\abstractname}{}

\begin{abstract}

\bf Abstract \rm This study presents a statistical analysis of 233 geomagnetic storms associated with coronal holes (CH) and predominantly driven by high-speed solar wind streams from them during the 25th solar cycle (2020-2026). Data from NASA CDAW, DONKI (NASA CCMC), WDC Kyoto, and STCE catalogs were utilized. It was shown that coronal holes became an important source of geomagnetic storms during the growth and decay phases of the cycle: during the peak activity (2023-2025), CH-storms accounted for more than 80$\% $ of all geomagnetic events. A lack of correlation between solar wind speed ($V_{\mathrm{SW}}$) and storm intensity (Dst) ($r \approx 0$) was established. The main drivers of geoeffectiveness were found to be the southward component of the interplanetary magnetic field ($B_z$) and the duration of impact. Seasonal modulation (Russell--McPherron effect) manifested in the concentration of 60.5$\%$ of storms during spring and autumn months. These findings are significant for long-term space weather forecasting.


\end{abstract}

\section{Introduction}

Geoeffective events represent a complex set of phenomena influencing
near-Earth space. They can cause notable effects on Earth, including magnetic
and radiation storms, auroras, disruptions in radio communications, navigation
issues, and risks for satellites and astronauts.

The main sources of such events include solar flares, whose geoeffectiveness
depends on their power and position on the solar disk, coronal mass ejections
(CMEs), eruptive prominences, and high-speed solar wind streams originating
from coronal holes (HSS).
In modern heliophysics, two main mechanisms for transferring energy from the
Sun to Earth's magnetosphere are distinguished \cite{yermolaev2010}. The first
scenario involves a solar flare, followed by CME that evolves into an
interplanetary CME (ICME) with a southward-directed Bz component, ultimately
triggering a geomagnetic storm. The second scenario involves a coronal hole,
which gives rise to a high-speed solar wind stream (HSS) and a subsequent
corotating interaction region (CIR) characterized by a southward Bz component,
also leading to a geomagnetic storm. This paper focuses primarily on the
second mechanism; it also evaluates the contribution of coronal holes as one
of the important sources of geomagnetic activity during the 25th solar cycle.

Coronal holes (CHs) are regions in the solar corona with an open magnetic
field configuration, through which plasma flows unimpeded into interplanetary
space, forming HSSs. Unlike CMEs---which dominate during the solar cycle
maximum and trigger extreme, albeit short-lived, storms---CH-driven streams
sustain geomagnetic activity during the rising and declining phases of the
solar cycle \parencite{tsurutani2006,richardson2018}. The effectiveness of the
resulting geomagnetic impact is determined by a complex set of factors,
including solar wind parameters, the characteristics of the interplanetary
magnetic field, heliospheric conditions, and the orientation of the
geomagnetic field. \textcite{abunina2014} analyzed the geoeffectiveness of
coronal holes in 2011--2012 as a function of heliolatitude and polarity. The
authors concluded that trans-equatorial coronal holes are the most
geoeffective. CHs with negative polarity in northern latitudes and positive
polarity in southern latitudes exhibit a less pronounced, yet still
significant, impact.

During Solar Cycle~24 (2008--2019), no extremely strong geomagnetic storm was
recorded ($K_p = 9$, $A_p \approx 400$~nT), which is partly attributed to the
reduced geoeffectiveness of CMEs \parencite{stce_sc25}.
At the same time, high-speed streams from coronal holes remained a significant
source of geomagnetic activity. Solar Cycle~25, which began in December 2019,
has exceeded the NOAA predictions: the smoothed sunspot number reached
$\sim$161 in October 2024, while the monthly mean temporarily exceeded 200 in
August 2024, far surpassing the forecast value of 115
\parencite{noaa_sc25,stce_sc25}. This elevated level of solar activity makes
it timely to reassess the role of coronal holes in shaping geomagnetic
conditions.

The analysis of geomagnetic storms relies on a set of indices, including Dst,
$V_{\mathrm{SW}}$, $B_z$, $K_p$, and the NOAA G-class. Among these, the
minimum Dst value reached during an event is a principal quantitative measure
of storm intensity: it reflects the global intensification of the ring current
in the magnetosphere.

For coronal-hole-driven storms (CH-storms), the Dst minimum serves a dual
role. First, it provides a consistent criterion for event selection and
comparison: by applying a uniform Dst threshold (e.g.,
$\mathrm{Dst} \leq -30$~nT), we ensure a homogeneous sample of geoeffective
disturbances regardless of the underlying solar wind conditions. Second,
within this curated sample, the Dst index becomes the dependent variable for
statistical analysis, enabling the assessment of how solar wind drivers (such
as $V_{\mathrm{SW}}$ and $B_z$) influence the resulting geomagnetic response.

To implement this approach, a dedicated event dataset of coronal-hole-driven
disturbances during Solar Cycle~25 was constructed. The dataset includes key
parameters for each event: onset time, duration, $V_{\mathrm{max}}$,
$B_{\mathrm{max}}$, minimum $B_z$, minimum Dst, and maximum $K_p$. This
dataset forms the basis for a systematic evaluation of the relationship
between solar wind conditions and geomagnetic activity, with a focus on
isolating the contribution of coronal holes.

The aim of this work is to quantify the contribution of coronal holes to the
geomagnetic activity of Solar Cycle~25 using the Dst index and to identify
regularities with respect to the cycle phase, seasonal distribution (including
the Russell--McPherron effect), and solar wind parameters.

\section{Data and methods}

This section describes the data sources, selection criteria, and
classification scheme used to construct the event catalogue and to assess the
statistical relationship between solar wind drivers and geomagnetic response.
The methodology is designed to ensure unambiguous source attribution (CH vs
CME vs mixed) and to provide consistent metrics for quantifying storm
intensity and solar wind conditions.

\subsection{Data sources}

Coronal holes and coronal mass ejections were identified using the DONKI
database developed at the NASA Community Coordinated Modeling Center (CCMC)
\parencite{donki}. To verify the evolution of solar features and the
spatial characteristics of coronal holes, we additionally employed the Solar
Terrestrial Activity Report \parencite{solar_terrestrial}, which provides
independent estimates of coronal hole areas and locations. This two-source
approach enabled robust cross-validation of CH identification and reduced the
risk of misclassification.

The Dst index, a key metric for geomagnetic storm intensity, was obtained from
three complementary sources: the NASA CDAW Dst catalogue \parencite{cdaw_dst},
the World Data Center (WDC) Kyoto hourly Dst values \parencite{wdc_kyoto}, and
the NOAA SWPC geomagnetic storm catalogue \parencite{swpc_storms}. These
datasets provide consistent coverage of storm occurrence, timing, and
intensity, enabling robust cross-validation of event parameters.

Solar wind plasma and magnetic field parameters ($V_{\mathrm{SW}}$, $B_z$,
proton density $n$, and temperature $T$) were taken from the OMNIWeb dataset
\parencite{omniweb}, combining observations from Wind and ACE spacecraft at
1~AU. These data were used to identify high-speed streams and stream
interaction regions (CIRs), as well as to derive event-specific metrics: peak
solar wind speed ($V_{\mathrm{max}}$), maximum magnetic field
($B_{\mathrm{max}}$), and minimum $B_z$.

Summary statistics for Solar Cycle~25 were obtained from the STCE SC25
Tracking service \parencite{stce_sc25}. For selected events, additional
diagnostics were retrieved from the interactive database of IZMIRAN
\parencite{izmiran_db}.

All online resources were accessed on 2026-09-07 unless otherwise noted.

\subsection{Event selection and classification criteria}

CH-storms were selected based on two primary criteria: (1) a minimum storm
intensity threshold of $\mathrm{Dst} \leq -30$~nT, ensuring inclusion of at
least minor storm activity; (2) an association with a coronal hole, confirmed
either by direct identification in the DONKI catalogue
\parencite{donki} or by the presence of a high-speed solar wind stream
($V_{\mathrm{SW}} > 400$~km\,s$^{-1}$) and a corresponding stream interaction
region (CIR) in the OMNI data \parencite{omniweb}. Events driven primarily by
CMEs without a significant CH contribution were excluded, except for events
that entered the sample by their Dst minimum timing during the presence of a
CH-associated stream; such mixed events are retained and explicitly flagged.

For each event, onset time and duration, $V_{\max}$, $B_{\max}$, minimum $B_z$,
minimum Dst and maximum $K_p$ were recorded. These quantities were used to
characterize the solar wind driver and the resulting geomagnetic response, as
well as to evaluate the dependence of storm intensity on solar wind
conditions.

Storms were classified in two independent ways: by the NOAA G-scale, defined
by $K_p$ (G1: $K_p=5$; G2: $K_p=6$; G3: $K_p=7$; G4: $K_p=8$; G5: $K_p=9$;
events with $K_p \leq 4^+$ below G1), and by peak Dst intensity following the
conventional scheme \parencite{gonzalez1994,loewe1997}: weak
($-50 < \mathrm{Dst} \leq -30$~nT), moderate ($-100 < \mathrm{Dst} \leq -50$~nT),
intense ($\mathrm{Dst} \leq -100$~nT). Throughout the paper, G-classes refer
to the Kp-based NOAA scale, whereas weak/moderate/intense refer to Dst.

\subsection{Seasonal and statistical analysis}

To examine the seasonal distribution of CH-driven storms and to test for the
Russell--McPherron effect \parencite{russell1973,diego2021}, events were
distributed by month. This allowed for the identification of preferred seasons
for enhanced geomagnetic activity under different solar wind regimes.

Statistical analysis included the calculation of Pearson correlation
coefficients between key solar wind parameters ($V_{\mathrm{SW}}$, $B_z$) and
geomagnetic indices (Dst, $K_p$), as well as the determination of median and
mean values for each storm category. These metrics were used to quantify the
typical characteristics of CH-driven disturbances and to identify systematic
trends across the cycle phase and seasonal bins.

\section{Results}

\subsection{General statistics}

From September 2020 to July 2026, a total of 233~CH-associated storms with
$\mathrm{Dst} \leq -30$~nT were recorded. Of these, 174 reached moderate
intensity or higher ($\mathrm{Dst} \leq -50$~nT, 75\% of the sample), and 16
were classified as intense storms ($\mathrm{Dst} \leq -100$~nT) according to
the Dst-based intensity classification \parencite{gonzalez1994,loewe1997}.
Table~\ref{tab:ch-storms-by-year} presents the distribution of the
CH-associated storms by year and NOAA G-scale category, while
Table~\ref{tab:sample_stats} summarizes the key parameters of the sample.

The sample is dominated by weak and moderate activity: 80 events (34\%)
remained below G1, 97 reached G1 (42\%), 37 reached G2 (16\%), 14 reached G3
(6\%), and 5 events (2\%) reached G4. Thus, about 92\% of the CH storms fall
into the G2 or weaker categories, consistent with the established role of
coronal-hole streams as a source of moderate but prolonged geomagnetic
activity \parencite{reeve2021,andreeva2025}. All five G4 storms
(24~March 2023, $\mathrm{Dst} = -163$~nT; 28~June 2024, $-105$~nT;
17~September 2024, $-116$~nT; 16~April 2025, $-127$~nT; and 1~June 2025,
$-120$~nT) were mixed CH\,+\,CME events, confirming that the deepest storms
of the sample required a CME contribution superimposed on a CH-associated
stream. Notably, eight of the fourteen G3 storms occurred in 2025 alone ---
more than half of all G3 events in the cycle --- during the declining phase.

\begin{table}[!htbp]
\centering
\caption{Distribution of CH-associated geomagnetic storms by year and NOAA
G-class, solar cycle 25 (2020--2026).}
\label{tab:ch-storms-by-year}
\begin{tabular}{l r r r r r r}
\hline
Year & $<$G1 & G1 & G2 & G3 & G4 & Total \\
\hline
2020 &  7 &  3 &  1 &  0 & 0 &  11 \\
2021 & 18 & 10 &  3 &  1 & 0 &  32 \\
2022 &  8 & 21 &  6 &  0 & 0 &  35 \\
2023 &  8 & 18 &  5 &  2 & 1 &  34 \\
2024 & 13 &  9 &  6 &  2 & 2 &  32 \\
2025 &  4 & 21 &  8 &  8 & 2 &  43 \\
2026 & 22 & 15 &  8 &  1 & 0 &  46 \\
\hline
Total & 80 & 97 & 37 & 14 & 5 & 233 \\
\hline
\end{tabular}

\vspace{4pt}
\begin{minipage}{0.92\linewidth}
\footnotesize
\textbf{Notes:} G-classes follow the NOAA geomagnetic storm scale, defined by the
planetary $K_p$ index. $K_p$ values for 2026 events are provisional (predicted);
the 2026 data are therefore preliminary.
\end{minipage}
\end{table}

\begin{table}[!htbp]
\centering
\caption{Summary statistics of the CH-associated storm sample \\
(Solar Cycle~25, 24.09.2020--13.07.2026, $N = 233$).}
\label{tab:sample_stats}
\begin{tabular}{l r}
\hline
Parameter & Value \\
\hline
Number of events & 233 \\
Dst range (nT) & $-32$ \ldots $-163$ \\
Median Dst (nT) & $-60$ \\
$V_{\mathrm{SW}}$ range (km\,s$^{-1}$) & 386 \ldots 1349 \\
Median $V_{\mathrm{SW}}$ (km\,s$^{-1}$) & 580 \\
$B_z$ range (nT) & $-27.1$ \ldots $-4.4$ \\
$B_z$ coverage & 141 of 233 events \\
\hline
G4 storms & 5 \\
G3 storms & 14 \\
G2 storms & 37 \\
G1 storms & 97 \\
Below G1 & 80 \\
\hline
Storm classes (weak / moderate / intense) & 59 / 158 / 16 \\
Cycle phases (min / growth / max / decline) & 11 / 101 / 32 / 89 \\
\hline
\end{tabular}
\end{table}

\subsection{Phase portrait in the $V_{\mathrm{SW}}$--Dst plane}

\begin{figure}[!htbp]
\centering
\includegraphics[width=1.0\linewidth]{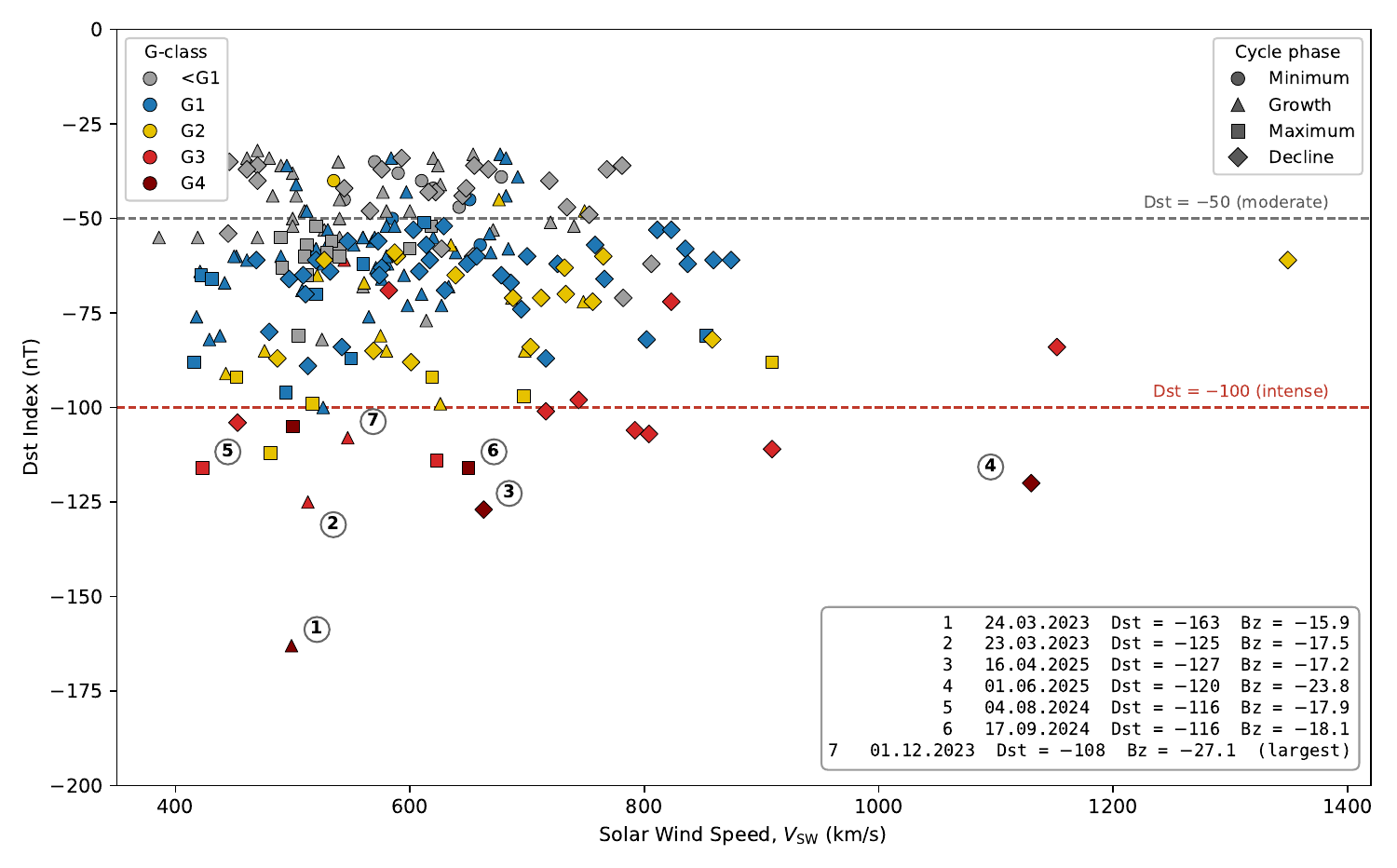}
\caption{Phase portrait of CH-associated storms in the Dst--$V_{\mathrm{SW}}$
plane for Solar Cycle~25 (2020--2026). Marker color encodes the NOAA storm
class; marker shape indicates the solar cycle phase. The deepest events
($\mathrm{Dst} \leq -116$~nT) are labeled (see Table~\ref{tab:key_storms});
the $-116$~nT level is shared by two events (04.08.2024 and 17.09.2024).
The event of 01.12.2023 (Dst $= -108$~nT) is additionally marked as having the
largest southward IMF component in the sample ($B_z = -27.1$~nT). Dashed lines
mark the Dst thresholds of $-50$~nT (moderate) and $-100$~nT (intense)
separating the Dst-based intensity categories
\parencite{gonzalez1994,loewe1997}.}
\label{fig:1}
\end{figure}

The phase portrait of CH-driven storms in the $V_{\mathrm{SW}}$--Dst plane
(Figure~\ref{fig:1}) reveals a lack of a direct linear relationship
between solar wind speed and storm intensity.

For the full sample of 233~events,
the Pearson correlation coefficient is weak and statistically insignificant
($r = -0.11$, $n = 233$; the critical value at the 0.05 level is $\approx 0.13$),
indicating that higher $V_{\mathrm{SW}}$ does not necessarily imply a deeper Dst
minimum. The scatter of points across the plane spans a wide range of solar
wind speeds, from $386$~km\,s$^{-1}$ to $1349$~km\,s$^{-1}$, with no single
velocity range exclusively associated with the strongest storms.

Despite the absence of a strong correlation, the distribution exhibits a
pronounced concentration of events in a well-defined region of the parameter
space. Approximately 59\% of CH storms fall within the domain
$350 \leq V_{\mathrm{SW}} \leq 650$~km\,s$^{-1}$ and
$-80 \leq \mathrm{Dst} \leq -30$~nT; extending the domain to
$300$--$700$~km\,s$^{-1}$ and $-100$ \ldots $-30$~nT captures about 79\% of the
sample. This region corresponds to the typical characteristics of coronal
hole-driven disturbances during Solar Cycle~25 and reflects the dominance of
moderate-speed high-speed streams and their associated stream interaction
regions (CIRs) in driving recurrent geomagnetic activity.

The observed pattern is consistent with the established role of coronal holes as
a source of moderate but prolonged geomagnetic disturbances
\parencite{reeve2021,andreeva2025}. The lack of strong correlation with
$V_{\mathrm{SW}}$ alone underscores the importance of additional drivers, in
particular the southward component of the interplanetary magnetic field ($B_z$).
This is further supported by the fact that the deepest event in the sample
(the mixed CH+CME storm of 24~March 2023, $\mathrm{Dst} = -163$~nT) occurred at
a merely moderate speed ($V_{\mathrm{SW}} = 499$~km\,s$^{-1}$, below the sample
median of 580~km\,s$^{-1}$) and was driven by extended periods of negative $B_z$,
highlighting the necessity of considering both plasma and magnetic field
parameters when characterizing the geoeffectiveness of solar wind structures.

The most geoeffective CH-associated events (Table~\ref{tab:key_storms})
reveal a consistent pattern: in every case the storm depth was set by a
southward IMF carried by a CME, while the coronal hole provided the background
flow and, in several events, a prolonged stream contribution.
The 24~March 2023 storm ($\mathrm{Dst} = -163$~nT, the deepest in the sample)
coincides with the onset of the geomagnetic disturbance interval associated with
the southern coronal hole CH1136, whose stream reached a peak of
658~km\,s$^{-1}$ yet produced only moderate activity on its own
(Kp$_{\max}=5$; \parencite{solar_terrestrial});
the storm depth was therefore driven by a stealth CME whose southward field
supplied $B_z = -15.9$~nT. The nearby event of 23~March 2023
($\mathrm{Dst} = -125$~nT) shares this pattern: a southern CH stream
($V_{\mathrm{SW}} = 513$~km\,s$^{-1}$) combined with a CME-related southward
field ($B_z = -17.5$~nT) produced the third-deepest storm of the sample.
The 1~December 2023 event ($\mathrm{Dst} = -108$~nT) was produced primarily by a
CME associated with the M9.8 flare of 28~November 2023, with an additional
contribution from the high-speed stream of CH1190; it reached its depth at a
moderate speed (547~km\,s$^{-1}$) owing to the record southward $B_z$ in our
sample ($-27.1$~nT). The 4~August 2024 storm formally entered the CH-associated
sample by its minimum timing but its driver was a slow ICME from the M8.2 flare of
1~August 2024, with the CH1233 contribution negligible: at
$V_{\mathrm{SW}} = 423$~km\,s$^{-1}$, the sustained southward field
($B_z = -17.9$~nT over many hours) was sufficient to reach
$\mathrm{Dst} = -116$~nT, illustrating that storm depth is governed by the
time-integrated southward field rather than by solar wind speed alone.
The 17~September 2024 storm ($\mathrm{Dst} = -116$~nT) resulted from a CME
associated with the X4.5 flare of 14~September arriving against a polar coronal
hole stream ($V_{\mathrm{SW}} = 650$~km\,s$^{-1}$, $B_z = -18.1$~nT).
The 16~April 2025 event ($\mathrm{Dst} = -127$~nT) was driven by a pair of
filament-eruption CMEs of 12--13~April arriving against the decaying high-speed
stream of CH1284. Finally, the 1~June 2025 storm ($\mathrm{Dst} = -120$~nT)
combines the recurrent trans-equatorial hole CH1298 with a fast CME (M8.1,
30~May), producing one of the highest speeds in the sample
($1130$~km\,s$^{-1}$) together with a strong southward $B_z = -23.8$~nT.
Thus, none of the deepest events was driven by a coronal hole alone. By
contrast, the deepest purely CH-stream-driven event of the sample occurred on
12~September 2024 ($\mathrm{Dst} = -114$~nT, G3), with the second-largest
southward IMF component ($B_z = -26.3$~nT); pure CH high-speed streams in Solar
Cycle~25 therefore appear to bound storm depth at the moderate-to-high (G2--G3)
level, whereas the deepest CH-associated storms result from the superposition of
CH streams and CMEs.

A complementary illustration is provided by the contrast between the storm of
1~December 2023, which reached $\mathrm{Dst} = -108$~nT owing to the largest
southward IMF component in the sample ($B_z = -27.1$~nT) despite a moderate
solar wind speed ($V_{\mathrm{SW}} = 547$~km\,s$^{-1}$), and the fastest event
of the cycle (9~March 2025, $V_{\mathrm{SW}} = 1349$~km\,s$^{-1}$), which
produced only a G2-level disturbance ($\mathrm{Dst} = -61$~nT;
Figure~\ref{fig:1}). This contrast further illustrates that the
southward IMF component, rather than the bulk flow speed, controls storm
intensity.

Notably, the deepest event (24~March 2023) occurred during the growth phase,
before the maximum of Solar Cycle~25, which peaked in October 2024
\parencite{sidc2025}.

\begin{table}[!htbp]
\centering
\caption{Key CH-associated storms of Solar Cycle~25 (2020--2026) annotated
in the phase portrait of Figure~\ref{fig:1}: the six deepest
events and the storm with the largest southward IMF component.}
\label{tab:key_storms}
\small
\setlength{\tabcolsep}{4pt}
\begin{tabular}{l r r r c c l}
\hline
Date & Dst & $V_{\mathrm{SW}}$ & $B_z$ & G & Phase & Driver \\
 & (nT) & (km\,s$^{-1}$) & (nT) & & & \\
\hline
24.03.2023 & $-163$ & 499  & $-15.9$ & G4 & Growth  & CH $+$ CME$^{a}$ \\
16.04.2025 & $-127$ & 663  & $-17.2$ & G4 & Decline & CH $+$ CME$^{b}$ \\
23.03.2023 & $-125$ & 513  & $-17.5$ & G3 & Growth  & CH $+$ CME$^{c}$ \\
01.06.2025 & $-120$ & 1130 & $-23.8$ & G4 & Decline & CH $+$ CME$^{d}$ \\
04.08.2024 & $-116$ & 423  & $-17.9$ & G3 & Maximum & ICME$^{e}$ \\
17.09.2024 & $-116$ & 650  & $-18.1$ & G4 & Maximum & CME $+$ CH$^{f}$ \\
01.12.2023 & $-108$ & 547  & $-27.1$ & G3 & Growth  & CME $+$ CH$^{g}$ \\
\hline
\end{tabular}

\vspace{4pt}
\begin{minipage}{0.92\linewidth}
\footnotesize
\textbf{Notes:} The table lists the six deepest CH-associated storms and the
event of 1~December 2023, which ranks tenth by minimum Dst but carries the
largest southward IMF component of the sample ($B_z = -27.1$~nT); all seven
events are annotated in Figure~\ref{fig:1}. The level
$\mathrm{Dst} = -116$~nT is shared by two events. Three deeper-than-$-108$~nT
purely CH-driven storms (12~September 2024, $-114$~nT, G3; 3~March 2024,
$-112$~nT, G2; 2~June 2025, $-111$~nT, G3) are not annotated.
$^{a}$CH1136 (southern; stream peak 658~km\,s$^{-1}$, Kp$_{\max}=5$) with a
stealth CME.
$^{b}$CH1284 decaying stream with a pair of filament-eruption CMEs
(12--13~April 2025).
$^{c}$Southern CH stream with a CME-related southward field.
$^{d}$CH52 HSS with a CME from the M8.1 flare of 30~May 2025.
$^{e}$Slow ICME from the M8.2 flare of 1~August 2024; the CH1233 contribution
was negligible.
$^{f}$CME from the X4.5 flare of 14~September 2024 arriving against a polar
CH stream.
$^{g}$CME from the M9.8 flare of 28~November 2023, superposed on the CH1190
high-speed stream.
Coronal hole (CH) identifications follow \parencite{solar_terrestrial};
solar wind and $B_z$ values are from OMNI \parencite{omniweb}; cycle phases
from \parencite{stce_sc25}. Of the five G4 storms in the sample
(Table~\ref{tab:ch-storms-by-year}), four are listed here; the fifth, the
mixed CH+CME storm of 28 June 2024 ($\mathrm{Dst} = -105$~nT, $K_p = 8^-$),
lies below the $\mathrm{Dst} \le -116$~nT selection threshold of this table.
\end{minipage}
\end{table}

\subsection{Seasonal dependence and the Russell--McPherron effect}

\begin{figure}[!htbp]
  \centering
  \includegraphics[width=0.85\textwidth]{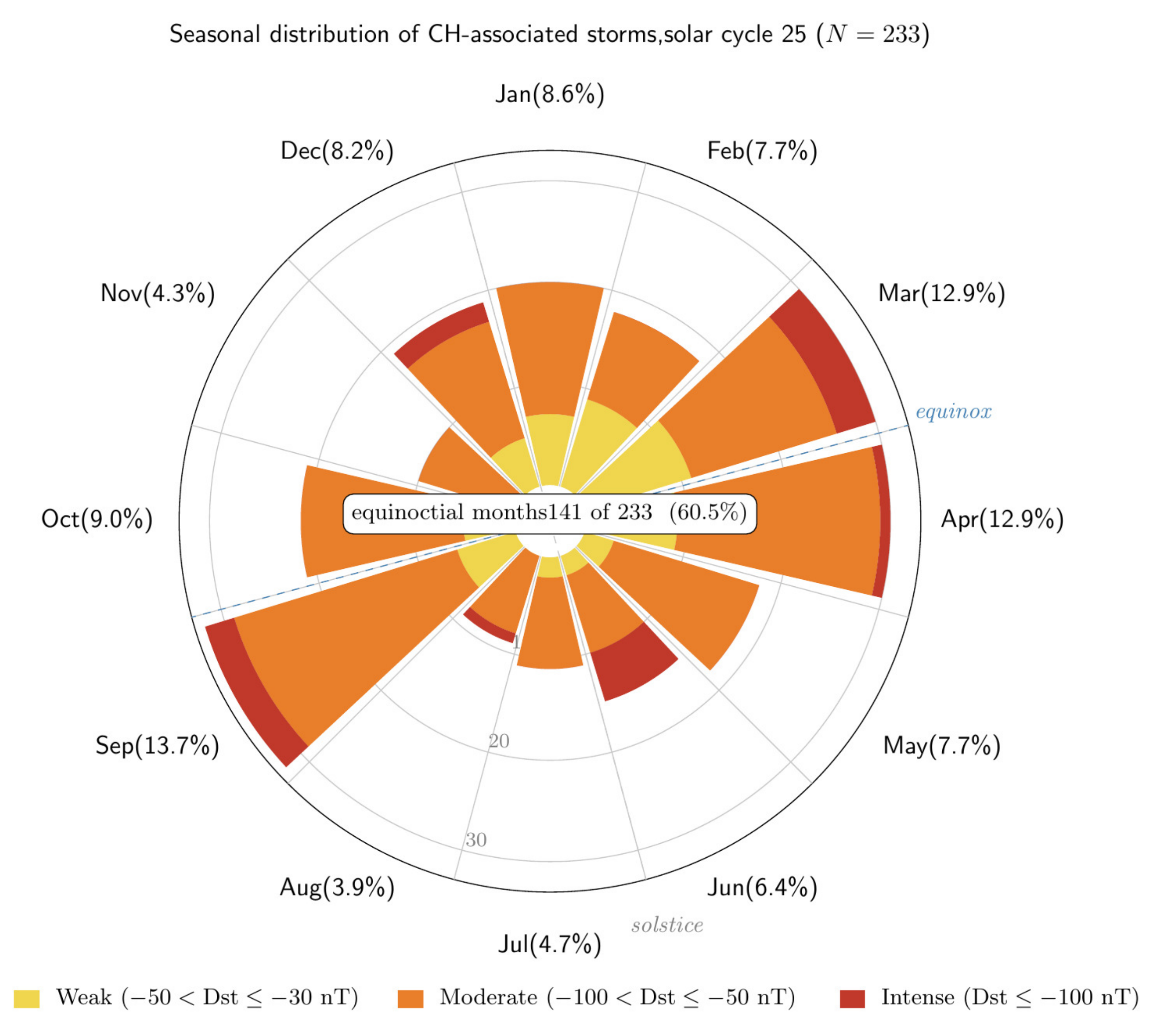}
  \caption{Seasonal distribution of CH-associated storms in Solar Cycle~25
    ($N=233$). Each sector corresponds to one calendar month; the sector
    height is proportional to the number of events, stacked by storm class
    according to the minimum $\mathrm{Dst}$ index: weak
    ($-50 < \mathrm{Dst} \le -30$~nT), moderate
    ($-100 < \mathrm{Dst} \le -50$~nT), and intense
    ($\mathrm{Dst} \le -100$~nT). Blue and gray dashed radii mark the
    equinoxes and solstices, respectively. The equinoctial months
    (March--May and September--November) accumulate 60.5\% of all events,
    consistent with the Russell--McPherron effect.}
  \label{fig:2}
\end{figure}

The monthly distribution of CH-driven storms (Figure~\ref{fig:2})
shows a pronounced equinoctial enhancement: the highest monthly counts fall
in September (13.7\%) and in March--April (12.9\% each), while the minimum
occurs in August (3.9\%); July contributes 4.7\%.

Spring and autumn together account for 60.5\% of all CH-driven storms,
in agreement with the Russell--McPherron effect
\parencite{russell1973}: near the equinoxes, the orientation of
the Earth's magnetic dipole with respect to the interplanetary magnetic
field becomes favourable for magnetic reconnection, which increases the
likelihood and efficiency of energy transfer into the magnetosphere.
This effect is particularly important for CH-driven storms, since their
geoeffectiveness is governed by the prolonged action of a steady flow
carrying a southward $B_z$, and near the equinoxes the probability of a
sustained negative $B_z$ is higher \parencite{nitti2023}.

According to \textcite{mourenas2021}, the seasonal modulation
of intense storms is expected to be more pronounced in odd-numbered solar
cycles than the modulation of moderate storms. Our data are consistent with
this picture: 10 of the 16 intense storms ($\mathrm{Dst} \le -100$~nT)
occurred in March--April or September--December. Notably, June alone
contributes five intense events --- the largest monthly count for this
class --- an outlier relative to the equinoctial pattern, reflecting
individual strong episodes rather than a systematic seasonal trend.

\section{Conclusions}

This work quantified the contribution of coronal holes to the geoeffective
events of Solar Cycle~25 based on a sample of 233 CH-associated storms
($\mathrm{Dst} \leq -30$~nT) recorded from September 2020 to July 2026.
The main conclusions are as follows.

\begin{enumerate}

  \item Coronal holes are an important source of geomagnetic storms in
  Solar Cycle~25, particularly during the growth and decay phases of the
  cycle: during the peak activity (2023--2025), CH-storms accounted for more
  than 80\% of all geomagnetic events. The sample is dominated by weak and
  moderate activity: about 92\% of the storms fall into the G2 category or
  weaker, consistent with the established role of coronal-hole streams as a
  source of moderate but prolonged geomagnetic disturbances.

  \item No linear relationship exists between solar wind speed and storm
  intensity ($r = -0.11$, statistically insignificant). The main drivers of
  geoeffectiveness are the southward component of the interplanetary magnetic
  field ($B_z$) and the duration of the impact: the fastest event of the
  cycle ($V_{\mathrm{SW}} = 1349$~km\,s$^{-1}$) produced only a G2-level
  disturbance, while the deepest storm ($\mathrm{Dst} = -163$~nT) occurred at
  a moderate speed of 499~km\,s$^{-1}$. A multivariate approach,
  $\mathrm{Dst} = f(V_{\mathrm{SW}},\,B_z,\,\Delta t)$, is required for
  forecasting.

  \item The deepest CH-associated storms of the cycle were all mixed
  CH\,+\,CME events, whereas purely CH-stream-driven disturbances appear to
  bound storm depth at the moderate-to-high (G2--G3) level. Mixed events
  therefore require separate treatment in space-weather forecasting models.

  \item A pronounced seasonal modulation (the Russell--McPherron effect) is
  confirmed: 60.5\% of all CH-driven storms occurred in the equinoctial
  months, with maxima in September (13.7\%) and March--April (12.9\% each)
  and a minimum in August (3.9\%). The distribution of intense storms is
  consistent with the expected stronger seasonal modulation in odd-numbered
  cycles.

  \item These results imply that long-term space-weather forecasting of
  CH-driven activity should rely on the southward IMF component and impact
  duration rather than on solar wind speed alone, and should take into
  account the predictable recurrence of coronal holes with a period of
  $\sim$27~days.

\end{enumerate}

\section*{Acknowledgements}

The author thanks the teams of the following organizations and services for
making their data publicly available: the NASA Community Coordinated Modeling
Center (CCMC) for the DONKI database, the NOAA Space Weather Prediction Center
for the geomagnetic storm catalogue (\texttt{swpc.noaa.gov}); the NASA Goddard
Space Flight Center for the CDAW
Dst catalogue and the OMNIWeb data service combining Wind and ACE observations
(\texttt{omniweb.gsfc.nasa.gov}); the World Data Center for Geomagnetism,
Kyoto University, for the hourly Dst values; the Solar-Terrestrial Centre of
Excellence (STCE) for the Solar Cycle~25 tracking statistics; the Pushkov
Institute of Terrestrial Magnetism, Ionosphere and Radio Wave Propagation
(IZMIRAN) for access to its interactive database; and the Solar Terrestrial
Activity Report (\texttt{solen.info}) for independent estimates of coronal
hole areas and locations.

This study was performed in the framework of the State assignment of the
Ministry of Science and Higher Education of the Russian Federation
(theme No.~122022400224-7).

\printbibliography[title={References}]

\end{document}